\documentclass[12pt,preprint]{aastex}

\def\kms  {km~s$^{-1}$}

\slugcomment{Draft: 18 Mar. 2006}

\shorttitle{} \shortauthors{Xu et al.}

\begin{document}

\title{Molecular outflows around high-mass young stellar objects}

\author{Y. Xu\altaffilmark{1,2}, Z. Q. Shen\altaffilmark{1}, J. Yang\altaffilmark{3},
X. W. Zheng\altaffilmark{4},  A. Miyazaki\altaffilmark{1}, K.
Sunada\altaffilmark{5}, H. J. Ma\altaffilmark{3}, J. J.
Li\altaffilmark{1,6}, J. X. Sun\altaffilmark{3} and C. C.
Pei\altaffilmark{3}}

\altaffiltext{1}{Shanghai Astronomical Observatory, Chinese
Academy of Sciences, Shanghai 20030, China; yxu@shao.ac.cn}
\altaffiltext{2}{Max-Planck-Institut f$\ddot{u}$r Radioastronomie,
Auf dem H$\ddot{u}$gel 69, 53121 Bonn, Germany;
xuye@mpifr-bonn.mpg.de} \altaffiltext{3}{Purple Mountain
Observatory, Nanjing, 210008, China} \altaffiltext{4}{Department
of Astronomy, Nanjing University Nanjing 210093, China}
\altaffiltext{5}{Nobeyama Radio Observatory, Nagano, 384-13,
Japan} \altaffiltext{6}{Graduate School of Chinese Academy of
Sciences, Beijing 100039, China}

\begin{abstract}
We present a study of molecular outflows using high-resolution
mapping of the CO (1-0) line emission toward eight relatively
nearby 6.7 GHz methanol masers, which are associated with massive
star forming regions. Outflows were detected in seven out of eight
sources, and five of them clearly show bipolar or multiple outflow
morphologies. These outflows have typical masses of a few solar
masses, momenta of tens of $M_{\odot}$ km s$^{-1}$, kinetic
energies of $\sim$ 10$^{45}$ ergs, and the mass entrainment rate
of a few $10^{-5} M_{\odot}$ yr$^{-1}$. They have significantly
more mass and kinetic energy than their low-mass counterparts. In
some of the sources, the massive outflow is obviously associated
with a particular massive star in the cluster, while in others the
origin remains uncertain.  The high detection rate of outflows
toward methanol masers suggests that the outflow phase of massive
protostars encompasses the methanol maser phase.
\end{abstract}


\section{INTRODUCTION}

Molecular outflows may be the main signature of the earliest stage
in massive star formation. Outflows have been investigated with CO
molecule lines by many authors (Bally \& Lada 1983; Lada 1985;
Snell et al. 1988; 1990), since Zuckerman et al. (1976) and Kwan
\& Scoville (1976) discovered them. However, most studies are
toward low-mass star-forming regions. Recently some systematic
observations toward massive star-forming regions have been carried
out (Shepherd \& Churchwell 1996; Zhang et al. 2001; 2005; Beuther
et al. 2002). It is difficult to determine the characteristics of
outflows associated with massive YSOs because there are relatively
fewer massive stars, they tend to form in clusters, and they are
often at distances greater than 1 kpc. Therefore, a moderate
resolution observation is critical to achieve the necessary degree
of kinematic and morphological details. Because of this, most of
the observations were carried out with the $^{12}$CO (2-1) line,
while only a few sources were studied by using the $^{12}$CO (1-0)
line with a high resolution. The excitation energy of the J=2
level is 16 K above the ground state, while it is only 5 K for the
J=1 level.  Thus the J=2 level (and higher) can better represent
the conditions thought to be in outflows, while the J=1 level can
trace the cooler, non-star-forming gas in the galactic plane.
Obviously, having two different transitions observed will be
helpful to our better understanding of the physical properties of
outflows.

In order to investigate the structure and dynamics of outflows
associated with massive star-forming regions on a finer scale, we
performed $^{12}$CO (1-0) observations of 6.7 GHz methanol masers
in a sample of massive star-forming regions (Xu et al. 2003) with
the Nobeyama 45-m telescope in Japan. It is the first systematic
study with a high resolution observation ($\sim 15''$) of the CO
(1-0) line. The 6.7 GHz methanol maser (a Class II methanol maser
transition) is often found to be associated with massive star
forming regions (Menten 1991; Walsh et al. 1998; van der Walt et
al. 2003). The selecting criteria are (1) declination $\delta$
$\geq -10^\circ$; and (2) distance d $\leq$ 1.5 kpc. There are
eight sources that satisfy these criteria (Table 1). These sources
have infrared luminosities between about $10^{3} L_{\odot}$ and
$10^{4} L_{\odot}$. Most of them have been observed previously in
CO by other single dish telescopes. Both IRAS 06053-0622 and
22543+6145 were also observed by interferometer (Minier et al.
2000). Our observations have a linear resolution of better than
$\sim$ 0.1 pc, about one tenth of the typical length of an
outflow. Thus they should be able to well determine the morphology
of outflows.

\section{OBSERVATIONS}
The observations were made in 2003 April at the $^{12}$CO {\it
J}=1--0 emission line (115.271202~GHz) with the 45-m radio
telescope of the Nobeyama Radio Observatory (NRO) in Japan. The
names, positions, and velocities of the observed 8 IRAS sources
are summarized in Table~1. We used the 25-beam array receiver
system (BEARS), which is the $5\times5$ focal-plane array SIS
mixer receiver working in the 80-115~GHz band (Sunada et al.
2000). Typical main beam efficiency and half power beam width
(HPBW) for each beam at 115~GHz are 0.45 and 15$''$, respectively.
The pointing accuracy was about $5''$ and was checked by observing
nearby SiO maser sources at 43~GHz every one hour. Since the beam
separation on the sky is $\sim 41''$, we observed the interpolated
two points between each beam using the multi ON-OFF position
switching technique, and then made the final map with the grid
spacing of $\sim14''$ in an area of $3'\times3'$ around the IRAS
sources. The OFF positions are about 1$^\circ$ away from each
source. The typical system temperature during the observations was
$T_{\rm sys}$(DSB) = 450~K. The intensity scale of the spectra was
calibrated by the chopper wheel method. The spectrometer is an
auto correlator of high resolution mode with a 32~MHz bandwidth
and a 37.8~kHz spectral resolution, corresponding to a velocity
resolution of $\sim0.1$~\kms\, at 115~GHz.

The intensities of the spectra were corrected using the scaling
factor provided by the NRO to calibrate the the sideband rejection
ratio of each BEARS element and to establish the absolute
intensity scale. The scaling factors of each beam are estimated by
comparing the peak intensities in S140 in the $^{12}$CO {\it
J}=1--0 line with those measured with a single sideband SIS
receiver (S100). Then the line intensity is reported in terms of
the corrected antenna temperature $T_{\rm R}^*$. The main beam
efficiency with S100 at 115~GHz is 0.40. The total integration
time of each position was about 6 minutes, which achieved a RMS
noise level of $\sim0.2$~K ($T_{\rm A}^*$) at the velocity
resolution of $\sim0.1$~\kms. The data were reduced by using the
NEWSTAR reduction package of the NRO. After subtracting linear
baselines, the data were smoothed to a resolution of $\sim \
$0.5~\kms\ to improve the signal to noise ratio.

\section{RESULTS AND DISCUSSION}
Molecular outflows were detected in all sources, except IRAS
22551+6139. The mapping results of the integrated CO line wing
emission are shown in Figure 1. Five sources (IRAS 00338+6312,
IRAS 06053$-$0622, IRAS 06056+2131, IRAS 20081+3122 and IRAS
22543+6145) show a clear bipolar or multiple outflow structure.
Except in IRAS 06053$-$0622, the methanol masers are located near
the center of the bipolar outflows. One source (IRAS 22272+6358)
appears to have a one-sided lobe. The ranges of red and blue wing
emission, the integrated flux at the position of maximum
integrated intensity, and the collimation factor (Lada 1985) are
listed in Table 2.

The outflow parameters, except for CO column density $N_{CO}$
which is derived from Snell et al. (1988), are estimated with the
method of Beuther et al. (2002). We assume that the gas is in
local thermodynamic equilibrium at a gas temperature of 30 K, and
a CO to H$_{2}$ abundance ratio of $10^{-4}$ (Snell et al. 1988).
The physical properties of outflows are summarized in Table 3. We
here are using the optically-thin assumption, and masses and
energetics in Table 3 are lower-limits to the true values. These
are about an order of magnitude less than the values with the
method of Beuther et al. (2002), where they corrected for optical
depth effects in their data. Typically, the molecular outflows
have masses of a few solar masses, momenta of tens of $M_{\odot}$
km s$^{-1}$, kinetic energies of $\sim$ 10$^{45}$ ergs, mass
entrainment rate of a few $10^{-5} M_{\odot}$ yr$^{-1}$,
mechanical force of a few $10^{-4} M_{\odot}$ km s$^{-1}$
yr$^{-1}$, and mechanical luminosity of about 1 $L_{\odot}$,
respectively. These results show that outflows from these massive
star forming regions have significantly more mass and kinetic
energy than those from low-mass star forming regions (Myers et al.
1988, Cabrit \& Bertout 1992, Bontemps et al. 1996).

\subsection{Comments on Individual Sources}
\noindent \textit{IRAS 00338+6312} \  IRAS 00338+6312 is located
near the core of the dark cloud L1287. The methanol maser is
coincident with the IRAS source. A bipolar outflow was first
detected by Snell et al. (1990) and confirmed by Yang et al.
(1991). The outflow is in the NE-SW direction and shows a clear
bipolarity. Evans et al. (1994) mapped CO (3-2) with a comparable
resolution of $20''$ to our CO (1-0) observations. They computed a
range of masses for the outflow using the optically-thin
assumption for the CO (3-2) data for the lower limit, and an
optical-depth correction for the upper limit. The lower limit
outflow mass derived by them is similar to that listed in Table 3
derived from the CO (1-0) data, as we would expect given the
optically-thin assumption for them. The IRAS source lies midway
between the center of the blue and red lobes. There is the FU Ori
star RNO1B located about 11$''$ south of the IRAS peak (Evans et
al. 1994).  However, the CO outflow appears to be driven not by
RNO1B, but by a source at or closer to the IRAS source. A radio
continuum source with a positive spectral index (Anglada et al.
1994), and a core observed with CS, HCN, HCO$^{+}$ and NH$_{3}$
(Yang et al. 1991; Walker \& Masheder 1997; Zinchenko et al. 1997)
are coincident with the IRAS source. All of these tracers appear
to originate from the same source that is driving the bipolar
outflow.\\

\noindent \textit{IRAS 06053-0622} \ There is a large offset
(18$''$, 3$''$) between the methanol maser and IRAS 06053-0622
which is slightly off a compact HII region (7$''$, 8$''$, Walsh et
al. 1998). The outflow in this source has been widely studied by
many authors. With a resolution of about 120$''$, Bally and Lada
(1983) firstly detected a large bipolar outflow with a maximal
radius of up to 2 pc. Observation with a higher resolution
(60$''$) shows that the outflow consists of a blue wing and two
separated red lobes (Meyers-Rice \& Lada 1991). Our observations
with a better resolution of 15$''$ reveal that the outflow roughly
consists of two pairs of bipolar lobes. The emission from the red
lobe is stronger than that from the blue lobe. The less-dominant
one with poor collimation factor is situated to the north. There
are a lot of infrared sources in this region (Aspin \& Walther
1990; Yao et al. 1997). IRAS 06053-0622 is surrounded by IRS 1, 2,
3 and a$_{i}$. The methanol maser coincides with CS, H$_{2}$CO,
HCN and H$_{2}$CO peaks which are associated with infrared source
IRS 3 (Giannakopoulou et al. 1997; Choi et al. 2000). Since the
IRAS source, IRS 3 and the compact HII region are far away from
the center of the dominant outflow, they are not likely to be the
exciting source of the outflow. IRS 6 lies midway between peaks of
blue and red lobes of the dominant one, and so probably it is this
infrared source that drives the outflow. No infrared source is
near the less-dominant
outflow.\\

\noindent \textit{IRAS 06055+2039} \ Although no outflow was
claimed previously, broad CO wing emission was seen by Shepherd \&
Churchwell. (1996). Our image shows an outflow. Although the
outflow appears in both red and blue wings, they seem not to be a
bipolar outflow. The two lobes have different orientations. The
red one extends in the SW-NE direction, while the blue one is
almost perpendicular to the red one. There is a large offset
(40$''$, 17$''$) between the methanol maser and the IRAS source.
The center of red wing is closer the methanol (15$''$, 0) than the
IRAS source (30$''$, 15$''$). The methanol maser is coincident
with a compact HII region, an H$_{2}$O maser, and CS and
far-infrared peaks (Lada et al. 1981; Shepherd \& Churchwell 1996;
Zinchenko et al. 1998; Tej et al. 2006). Therefore, they are the
same source as the exciting source of the red wing.  The center of
the blue wing coincides with a far-infrared core (Tej et al.
2006), indicating a possible exciting source of the outflow.
\\

\noindent \textit{IRAS 06056+2131} \ IRAS 06056+2131 is located in
a complex of molecular clouds. The methanol maser is nearly
coincident with the IRAS source and an UC HII region (Kurtz et al.
1994). Snell et al. (1988) detected a CO bipolar outflow, which is
in the northeast-southwest direction, and peaks of blue and red
wings are spatially separated. From figure 1 there are two bipolar
outflows, one (outflow 1 detected by Snell et al. 1988) on the
upper-left and another, outflow 2, newly detected by Xu et al.
(2004) on the right. Zhang et al. (2005) obtained a similar result
with a resolution of about 30$''$. The IRAS source is associated
with outflow 2. The blue-shifted emission is stronger than the red
one in outflow 1 while in contrast the red-shifted lobe is
dominant in outflow 2. Since the mapping area is limited, the blue
lobe of outflow 1 is not shown entirely. From low resolution
mapping of Snell et al. (1988), it should stretch beyond the map
boundary to the east. The red wing of outflow 2 extends from the
west to the east and connects outflow 2 to outflow 1. Therefore,
outflow 1 can also be regarded as a multiple outflow since its
blue wing connects the red wings of the two outflows. There are a
lot of objects in this cloud. Near-infrared polarimetric images
show that there are two infrared nebulae correspondingly
associated with the two outflows (Yao et al. 2000). The nebula
near the outflow 2 is associated with a cluster of stars. Outflow
2 is associated with a water source, ammonia and CS emission
peaks, and an UC HII (Verdes et al. 1989; Kompe et al. 1989; Kurtz
et al. 1994; Morata et al. 1997). These indicate that there are at
least a massive star and a cluster of low- and intermediate mass
stars associated with outflow 2. Due to its large momentum,
outflow 2 is most likely to be driven by the UC HII region.
Outflow 1 is associated with an infrared nebula and a CS peak.
This outflow may be driven by an embedded massive star/stars
associated with the nearby infrared nebula. \\

\noindent \textit{IRAS 20081+3122} \ Yang et al. (2002) detected a
high-velocity CO line wing in this region. Kumar et al. (2004)
detected a bipolar outflow using H$^{13}$CO$^{+}$ line. Our
observations of the CO line also show a bipolar outflow, though
the orientation is slightly different. The red lobe extends
roughly along the EW direction, and the blue one along the NS
direction, while the H$^{13}$CO$^{+}$ outflow aligns in the SW-NE
direction. The length of the former is nearly as two times as the
latter. The methanol maser is associated with IRAS 20081+3122,
NH$_{3}$ and mm continuum peaks and other masers, such as 44 GHz
methanol, 22 GHz H$_{2}$O and 6 GHz OH masers (Zheng et al. 1985;
Desmurs \& Baudry 1998; Kurtz et al. 2004; Kumar et al. 2004;
Kurtz \& Hofner 2005). Previous observations of cm continuum
emission reveal an UC HII region which coincides with the IRAS
source (Turner \& Matthews 1984; Zheng et al. 1985; Kurtz et al.
2004) and is almost located at the center of the outflow,
indicating a possible exciting source
of the outflow. \\

\noindent \textit{IRAS 22272+6358} \ An outflow was only detected
in blue lobe, as firstly discovered by Sugitani et al. (1989). The
methanol maser is coincident with IRAS 22272+6358 and a NH$_{3}$
core (Zinchenko et al. 1997). Because no cm continuum emission has
been detected, this source may be too young to have produced a
detectable compact HII region (Wilking et al. 1989). Near-infrared
observation shows that this source is a heavily extinguished
object (Ressler \& Shure 1991). The IRAS source is close to the
center of the outflow, indicating a possible exciting source
of the outflow. \\

\noindent \textit{IRAS 22543+6145}  \ Though outflows in this
region have been investigated by many authors (Bally \& Lada 1983;
Richardson et al. 1987; Narayanan \& Walker 1996), our observation
is the first single dish study at a resolution of 15$''$. Figure 1
clearly shows a bipolar structure with a relatively poor
collimation. The CO (3-2) maps show a similar configuration to our
result (Narayanan \& Walker 1996). We could not confirm it due to
the finite mapping area. The methanol maser is coincident with
IRAS 22543+6145, a mm emission peak and a CS core, and near the
edge of an NH$_{3}$ core (Torrelles et al. 1993; Narayanan \&
Walker 1996). The IRAS source lies midway between two compact HII
regions (Hughes \& Wouterloot 1984; Garay et al. 1996). The
separation (9$''$, 2$''$) between the IRAS source and the two HII
regions is less than the IRAS error ellipse (10$''$, 5$''$). We
could not determine whether the IRAS source is associated with
either HII region.  Since they all reside near the center of the
outflow, it is unclear which source drives the molecular outflow.
The IRAS source lies midway between the center of both blue and
red lobes, indicating that it is most likely to denote the
location of driving source of the outflow. There seems to be a
second red lobe at the South, which is likely to be the red wing
of another outflow, but it is uncertain due to the limited mapping
area.
\\

\noindent\textit{IRAS 22551+6139} \ No outflow was detected in
this region. Although there is a large infrared luminosity,
non-detection of radio continuum indicates that this source is
associated with an earlier stage of the star formation process,
before the embedded star has had a chance to form an UC compact
HII region. At the time, an outflow may be too small or/and weak
to detect.

\subsection{DISCUSSION}

It is believed that massive outflows occur at the early phases of
protostellar evolution and disappear when a compaction HII region
evolves sufficiently. Thus, the exciting source is either a
massive protostar or an UC/compact HII region. There is evidence
that methanol masers may trace a more narrowly constrained stage
in the life of a massive protostar. They appear to arise in hot
molecular cores but disappear sometime before the HII region phase
(Minier et al. 2002). The outflows were detected in seven out of
eight sources. Five methanol masers are associated with compact
HII regions and located near the center of the outflows,
indicating that they could denote the location of the driving
sources of these outflows. However, not all of the compact HII
regions are the exciting sources of the outflows. In IRAS
06053-0622 and IRAS 06056+2131 (outflow 1), obviously the exciting
sources are not the known compact HII regions. In IRAS 22543+6145,
it is unclear which source is the exciting source of the outflow,
maybe one of the two compact HII regions or the IRAS source. There
is no compact HII region at all in IRAS 22272+6358 and so the
exciting source is probably a massive protostar. Since massive
stars are usually born in clusters, massive outflows are most
likely to be excited by one of massive stars in a cluster, such as
IRAS 22543+6145. On the other hand, from Table 3 the dynamical
timescale of the outflows is about a few $10^{4}$ yr, while an UC
HII regions could live a few $10^{5}$ yr (Churchwell 1999). In our
sample, outflows were detected in seven out of eight sources. The
high detection rate of outflows toward methanol masers suggests
that the duration of the outflow phase encompasses the methanol
maser phase.

\section{SUMMARY}
We mapped the CO (1-0) emission toward eight massive star forming
regions using the 45-m radio telescope of the NRO in Japan. With
one exception all sources were detected to be associated with
massive outflows. Two are newly detected in the CO (1-0) line, and
some show much finer structures than before. Our observations have
detected an outflow in IRAS 06055+2039 for the first time; The
outflow configuration detected in the CO (1-0) line is slightly
different from that in the H$^{13}$CO$^{+}$ line in IRAS
20081+3122; Multiple outflows are definitely detected in IRAS
06053-0622 and IRAS 06056+2131, and most likely in IRAS 22543+6145
as well. These outflows have significantly more mass and kinetic
energy than their low-mass counterparts. Some massive outflows are
obviously associated with a particular massive star in the
cluster. This is consistent with the high detection rate but the
short dynamical timescale of the outflows.

We are grateful to the NRO staff for the excellent support. We
thank the anonymous referee for many useful suggestions and
comments, which greatly improved this paper. We thank K. Tatematsu
for the help of the data reduction. This work is supported in part
by the National Science Foundation of China under grants 10133020,
10373025 and 10573029. Z.-Q. Shen acknowledges support by the
One-Hundred-Talent Program of the Chinese Academy of Sciences.

\begin{table*}
\begin{flushleft}
\normalsize Table 1: List of Objects: Columns 1 and 2 are source
names. Columns 3 and 4 list the positions of sources. $V_{LSR}$
(column 5) is the velocity of the line peak. Columns 6 and 7 are
distances and infrared luminosities derived from IRAS flux
densities (Casoli et al. 1986).
\\[0.05mm]
\end{flushleft}
         \label{Tabiras}
      \[
           \begin{array} {clccccccccccc}
             \hline \hline
      \noalign{\smallskip}

{\parbox[t]{25mm}{\centering IRAS}}& {\parbox[t]{25mm}{\centering
Name}}&
{\parbox[t]{20mm}{\centering R.A.(2000)\\
\mbox{$\mathrm{(^h\;\;\;^m\;\;\;^s)}$}}}&
{\parbox[t]{20mm}{\centering DEC(2000) \\
\mbox{$(^\circ\;\;\;'\;\;\;'')$}}}& {\parbox[t]{20mm}{\centering
$V_{LSR}$\\ \mbox{\scriptsize (km s$^{-1}$)}}}&
{\parbox[t]{8mm}{\centering $D$\\ \mbox{\scriptsize (kpc)}}}&
{\parbox[t]{8mm}{\centering $L_{FIR}$\\ \mbox{\scriptsize
($10^{3}L_{\odot}$)}}}&

\\
       \noalign{\smallskip}
\hline
      \noalign{\smallskip}
\small \rm
  00338+6312 &$L1287$ &00\;36\;47.5 & 63\;29\;02 & -18.3 & 0.9 &0.8         \\
  06053-0622 &$Mon R2$ &06\;07\;48.0 & -06\;22\;57& 9.5   & 0.8 &11.0        \\
  06055+2039 &$S252A$ &06\;08\;35.5 & 20\;38\;59 & 8.6   & 1.5 &6.0         \\
  06056+2131 &$AFGL \ 6366S$ &06\;08\;41.2 & 21\;31\;04 & 2.5   & 1.5 &10.7        \\
  20081+3122 &$ON \ 1$ &20\;10\;09.1 & 31\;31\;37 & 11.1  & 1.4 &7.5         \\
  22272+6358 &$L1206$ &22\;28\;52.2 & 64\;13\;43 & -8.9  & 0.9 &0.8         \\
  22543+6145 &$Cep \ A \ East$  &22\;56\;19.1 & 62\;01\;57 & -11.7 & 0.7 &15.3        \\
  22551+6139 & &22\;57\;11.2 & 61\;56\;03 & -10.7 & 0.6 &7.8         \\

      \noalign{\smallskip}
\hline

         \end{array}
      \]
   \end{table*}

\begin{table*}
\begin{flushleft}
\normalsize Table 2: Parameters of outflows: Columns (2, 3) and
(4, 5) list the LSR velocity range $\Delta v$, and the peak
integrated wing emission $S$ for the blue- and red-shifted wings,
respectively. The last column is the collimation factor $f_{c}$.
$\Delta v$ is measured at $T_{A}^{*}$ = 0.2 K level (2 $\sigma$)
and included low-velocity outflows (Shepherd \& Churchwell 1996).
\\[0.05mm]
\end{flushleft}
         \label{Tabiras}
      \[
           \begin{array} {ccccccccccccc}
             \hline \hline
      \noalign{\smallskip}

{\parbox[t]{25mm}{\centering IRAS \\Source}}&
{\parbox[t]{15mm}{\centering $\Delta v_{b}$\\
\mbox{(km s$^{-1}$)}}}&
{\parbox[t]{15mm}{\centering $\Delta v_{r}$ \\
\mbox{(km s$^{-1}$)}}}& {\parbox[t]{15mm}{\centering $S_{b}$\\
\mbox{\scriptsize (K km s$^{-1}$)}}}& {\parbox[t]{15mm}{\centering
$S_{r}$\\ \mbox{\scriptsize (K km s$^{-1}$)}}}&
{\parbox[t]{15mm}{\centering $f_{c}$\\
\mbox{\scriptsize }}}&
\\
       \noalign{\smallskip}
\hline
      \noalign{\smallskip}
\small \rm
  00338+6312  &(-30, -19)  &(-14, -7)     &36.2   &31.2  &2.2 &  \\
  06053-0622a &(0, 7)      &(15, 23)      &22.4   &28.0  &1.7 &  \\
  06053-0622b &(0, 7)      &(15, 23)      &37.9   &56.2  &2.3 &  \\
  06055+2039  &(0, 5.5)    &(12, 17)      &29.6   &36.1  &3.2 &  \\
  06056+2131a &(-10, 0)    &(5, 15)       &24.7   &12.4  &1.7 &  \\
  06056+2131b &(-10, 0)    &(5, 15)       &17.0   &25.6  &1.8 &  \\
  20081+3122  &(-5, 7.5)   &(15,20)       &17.5   &17.9  &2.7 &  \\
  22272+6358  &(-18, -11)  &              &8.0    &      &2.6 &  \\
  22543+6145  &(-28, -14.5)&(-6.5, 6)     &35.8   &28.3  &1.6 &  \\

      \noalign{\smallskip}
\hline

         \end{array}
      \]
   \end{table*}

\begin{center}
\begin{table*}
\begin{flushleft}
\normalsize Table 3: Results of Outflows: Columns 2 and 3 give
H$_{2}$ column densities $N$ [$10^{20}$ cm$^{-2}$] in both the
blue- and red-shifted outflow lobes with the corresponding lobe
masses in $M_{\odot}$ listed in columns 4 and 5. Column 6 is the
total mass ${M}_{out}$ [$M_{\odot}$]. Columns 7 and 8 give the
momentum $p$ [$M_{\odot}$ km s$^{-1}$], and the energy $E$
[$10^{45}$ erg]. The last five columns are the size [pc], the
characteristic time scale $t$ [$10^{4}$ yr], the mass entrainment
rate of the molecular outflow $\dot{M}_{out}$ [$10^{-5}
M_{\odot}$yr$^{-1}$], the mechanical force $F_{m}$ [$10^{-4}
M_{\odot}$ km s$^{-1}$ yr$^{-1}$], and the
mechanical luminosity $L_{m}$ [$L_{\odot}$], respectively.\\[0.05mm]
\end{flushleft}
\small
         \label{Tabiras}
      \[
           \begin{array} {clccrccccccccc}
             \hline \hline
      \noalign{\smallskip}

{\parbox[t]{10mm}{\centering Source \\ (1) }}&
{\parbox[t]{8mm}{\centering $N_{b}$\\ (2) }}
&{\parbox[t]{7mm}{\centering $N_{r}$ \\ (3) }}&
{\parbox[t]{5mm}{\centering $M_{b}$\\ (4) }}&
{\parbox[t]{5mm}{\centering $M_{r}$ \\ (5)
}}&{\parbox[t]{5mm}{\centering $M_{out}$\\ (6)
}}&{\parbox[t]{6mm}{\centering $p$\\  (7)
}}&{\parbox[t]{6mm}{\centering $E$\\  (8) }}&
{\parbox[t]{6mm}{\centering size\\  (9) }}&
{\parbox[t]{6mm}{\centering $t$\\  (10) }}&
{\parbox[t]{6mm}{\centering $\dot{M}_{out}$\\ (11) }}&
{\parbox[t]{8mm}{\centering $F_{m}$\\ (12 )}} &
{\parbox[t]{12mm}{\centering $L_{m}$ \\ (13) }}

\\
       \noalign{\smallskip}
\hline
      \noalign{\smallskip}
 00338+6312  &2.9&2.2&0.7&0.7&1.4&16.0&1.8 &0.37&3.1 &4.5 &5.1 &0.49   \\
 06053-0622a &2.6&3.1&0.4&0.6&1.0&11.0&1.4 &0.26&2.2 &4.5 &5.3 &0.53   \\
 06053-0622b &3.4&5.3&1.1&2.4&3.5&43.0&5.4 &0.34&2.9 &12.0&15.0&1.50   \\
 06055+2039  &2.8&3.3&3.3&3.4&6.7&60.0&5.5 &0.80&8.6 &7.7 &6.9 &0.51   \\
 06056+2131a &1.8&1.7&1.5&1.0&2.5&31.0&3.9 &0.41&3.2 &7.8 &9.8 &1.00   \\
 06056+2131b &1.6&2.5&0.9&2.2&3.1&39.0&4.9 &0.68&5.3 &6.0 &7.5 &0.77   \\
 20081+3122  &1.5&1.6&0.8&1.5&2.3&26.0&3.3 &0.73&5.7 &4.1 &4.6 &0.47  \\
 22272+6358  &0.5&   &0.3&   &0.3&2.5 &0.2 &0.49&5.2 &0.5 &0.5 &0.04  \\
 22543+6145  &3.1&2.7&0.5&0.3&0.8&14.0&2.4 &0.30&1.7 &4.7 &8.1 &1.10   \\

      \noalign{\smallskip}

\hline \hline

         \end{array}
      \]
\small

\normalsize

   \end{table*}

\end{center}

\begin{figure}
\epsscale{.65} \plotone{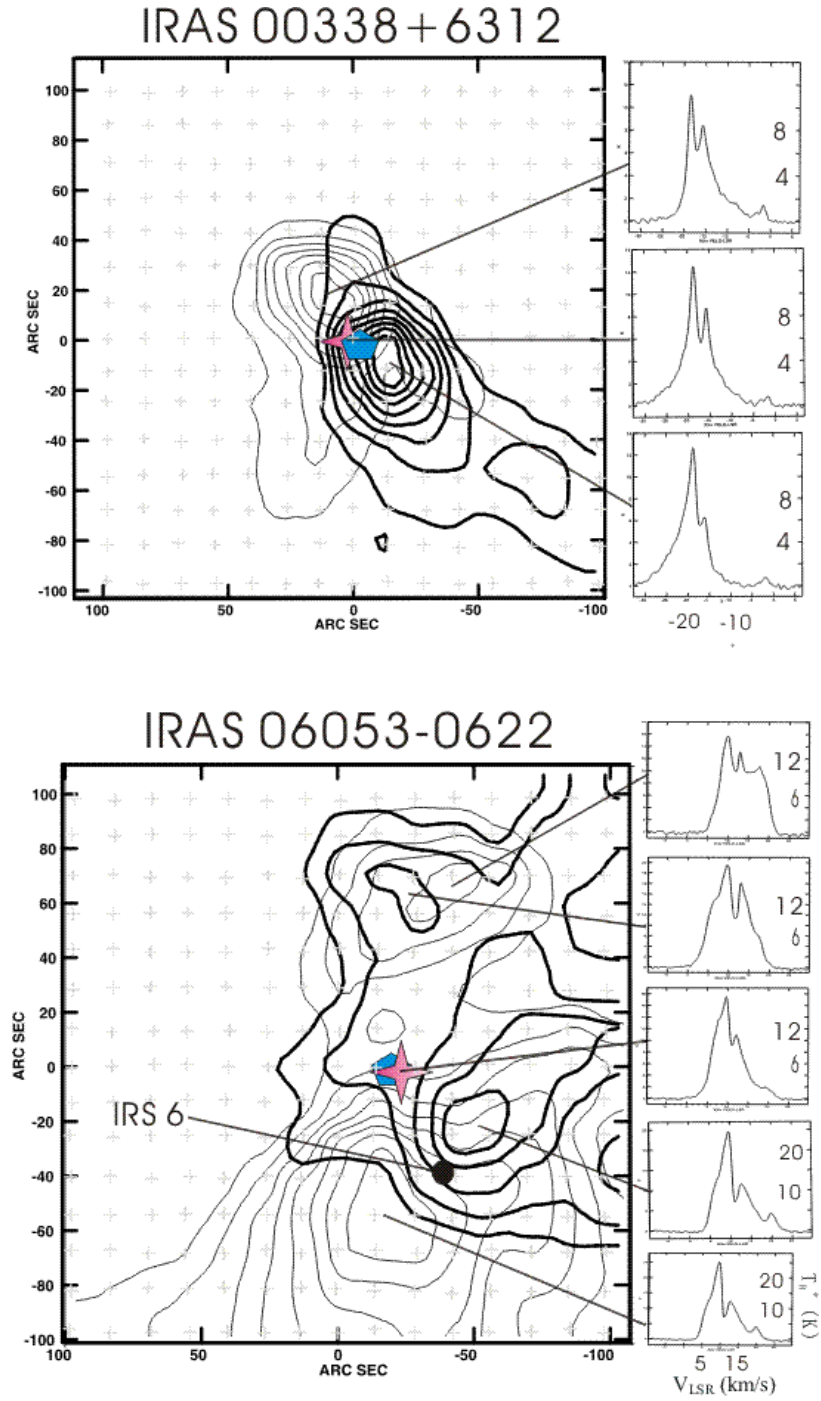} \caption{Maps of the integrated
wing emission. The 6.7 GHz Methanol maser emission is at the
origin (0,0). Red-shifted emission is indicated by thin contours,
while blue-shifted emission by thick contours. Contour levels are
20 to 90\% by steps of 10\% of the peak integrated wing intensity.
The star and pentagons symbols denote the IRAS source and HII
region, respectively. Small crosses are the observed points. The
spectra are plotted from the positions of the IRAS source, blue
and red peaks, respectively.}
\end{figure}
\begin{figure}
\epsscale{.85} \plotone{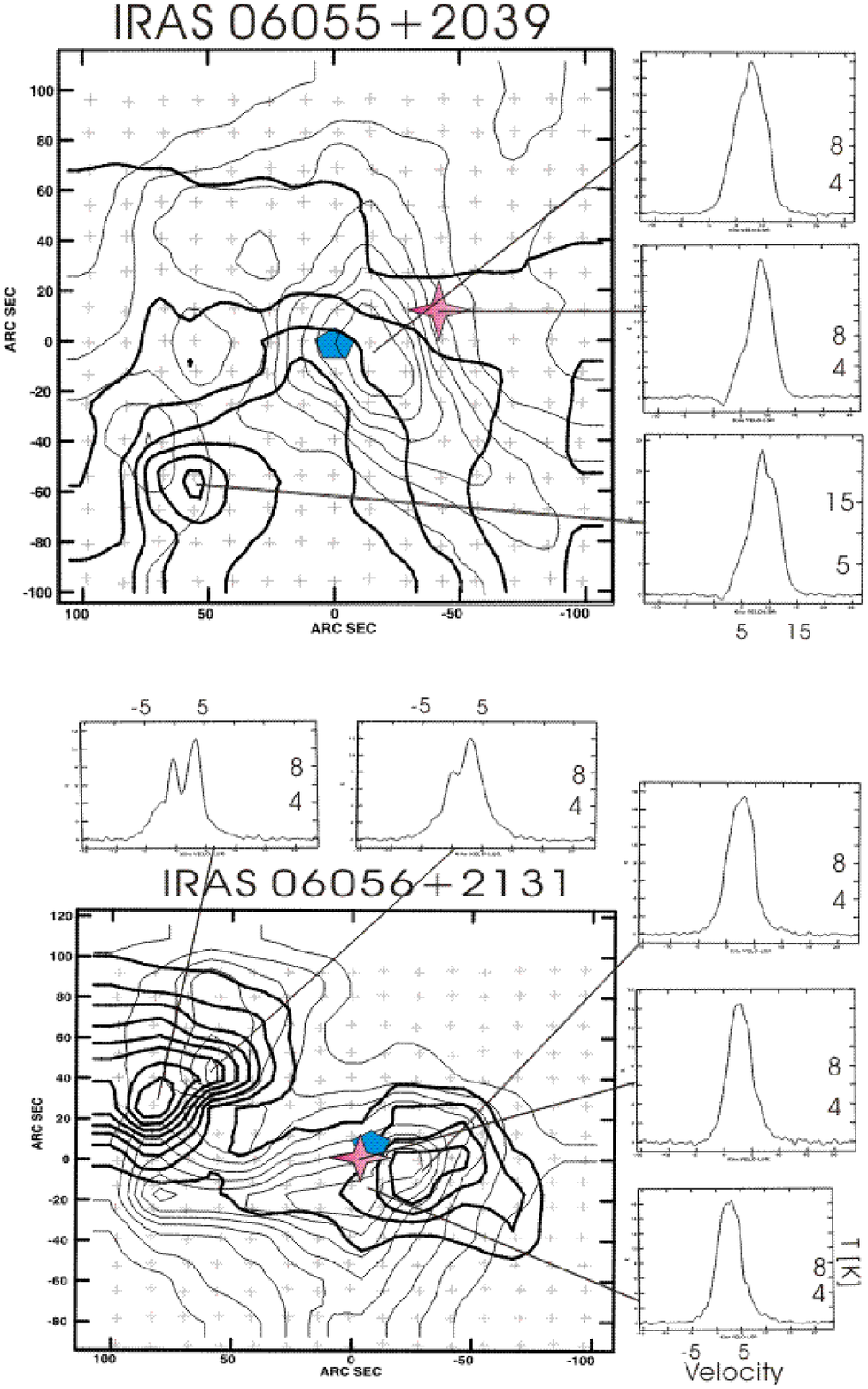} 
\end{figure}
\begin{figure}
\epsscale{.85} \plotone{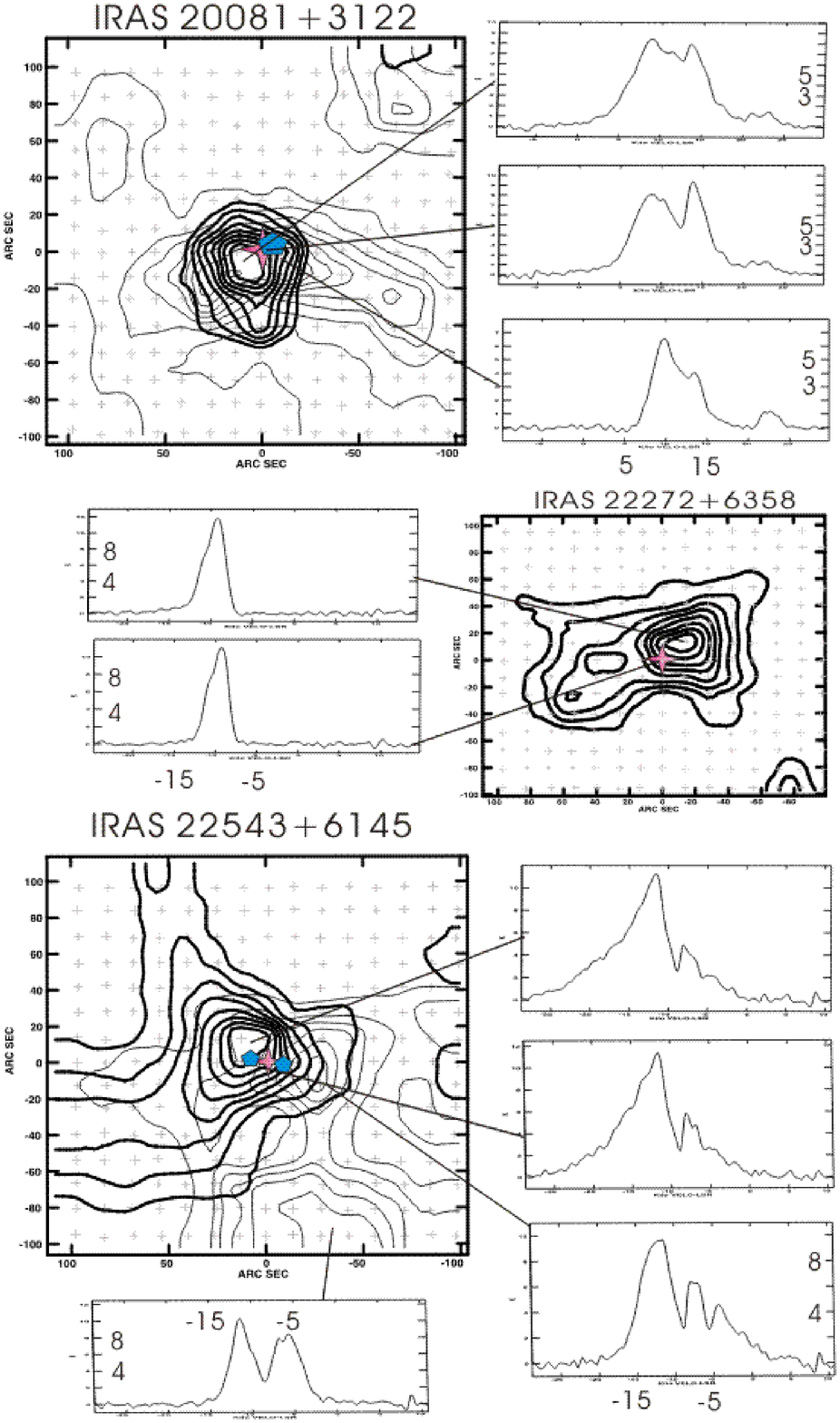} 
\end{figure}


\begin{thebibliography}{00}
{\small

\bibitem{} Anglada, G., Rodriguez, L. F., Girart, J. M., et al. 1994, \apj, 420, L91
\bibitem{} Aspin, C., \& Walther, D. M. 1990, A\&A, 235, 387
\bibitem{} Bally, J., \& Lada, C. J. 1983, \apj, 265, 824
\bibitem{} Beuther, H., Schilke, P., Sridharan, T.K. et al. 2002, A\&A, 383, 892
\bibitem{} Bontemps, S., Andr$\acute{e}$, P., Terebey, S., \& Cabrit, S. 1996, A\&A, 311, 858
\bibitem{} Cabrit, S., \& Bertout, C. 1992, A\&A, 261, 274
\bibitem{} Casoli, F., Dupraz, C., Gerin, M., Combes, F., \& Boulanger, F. 1986, A\&A, 169, 281
\bibitem{} Churchwell, E, 1999, The Origin of Stars and Planetary Systems.
Edited by Charles J. Lada and Nikolaos D. Kylafis. Kluwer Academic
Publishers, 1999, p. 515
\bibitem{} Choi, M., Evans II, N. J., Tafalla, M., \& Bachiller, R. 2000, \apj, 538, 738
\bibitem{} Desmurs, J. E., \& Baudry, A. 1998, A\&A, 340, 521
\bibitem{} Evans, N. J., Stephen, B., Russell, M. L., Lee, H., \& Scott, K. 1994, \apj, 424, 973
\bibitem{} Garay, G., Ramirez, S., Rodriguez, Luis F. et al. 1996, \apj, 496, 193
\bibitem{} Giannakopoulou, J., Mitchell, G. F., Hasegawa, T. I. et al. 1997, \apj, 487, 346
\bibitem{} Hughes V. A., \& Wouterloot J. G. A. 1984, \apj, 276, 204
\bibitem{} Kompe, C., Baudry, A., Joncas, G., \& Wouterloot, J. G. A. 1989 \aap, 221, 295
\bibitem{} Kumar, M. S. N., Tafalla, M., \& Bachiller, R. 2004, A\&A, 426, 195
\bibitem{} Kurtz, S., Churchwell, E., \& Wood, D. O. S. 1994, \apjs, 91, 659
\bibitem{} Kurtz, S., Hofner, P., \& $\acute{A}$lvarez, C. V. 2004, \apjs, 155, 149
\bibitem{} Kurtz, S., \& Hofner, P. 2005, \aj, 130, 711
\bibitem{} Kwan, J., \& Scoville, N. 1976, \apj, 210, L39
\bibitem{} Lada, C. J. 1985, ARA\&A, 23, 267
\bibitem{} Lada, C. J., Blitz, L., Reid, M. J., \& Moran, J. M. 1981, \apj,
243, 769
\bibitem{} Menten, K. 1991, \apj, 380, L75
\bibitem{} Meyers-Rice, B. A., \& Lada, C. J. 1991, \apj, 368, 445
\bibitem{} Morata, O., Estalella, R., Lopez, R., \& Planesas, P. et al 1997 \mnras, 292, 120
\bibitem{} Myers, P. C., Heyer, M., Snell, R. L., \& Goldsmith, P. F. 1988, \apj, 324, 907
\bibitem{} Minier, V., Booth, R. S., \& Conway, J. E., 2000, A\&A, 362, 1093
\bibitem{} Minier, V., Booth, R. S., Burton, M. G., \& Pestalozzi, M. R., 2002,
Proceedings of the 6th European VLBI Network Symposium Ros, E.,
Porcas, R.W., Lobanov, A.P., \& Zensus, J.A. (eds.) June 25th-28th
2002, Bonn, Germany, p. 205
\bibitem{} Narayanan, G., \& Walker, C. K. 1996, \apj, 466, 844
\bibitem{} Ressler, M. E., \& Shure, M. 1991, \aj, 102, 1398
\bibitem{} Richardson, K. J., White, Glen J., Avery, L. W., \& Woodsworth, A. W. 1987, A\&A, 174, 197
\bibitem{} Shepherd D.S., \& Churchwell, E. 1996, \apj, 457,267
\bibitem{} Snell, R. L., Huang, Y.-L., Dickman, R. L., \& Claussen, M. J. 1988, \apj, 325, 853
\bibitem{} Snell, R. L., Huang, Y.-L., Dickman, R. L., \& Claussen, M. J. 1990, \apj, 352, 139
\bibitem{} Sugitani, K., Fukui, Y., Mizuni, A., \& Ohashi, N. 1989, \apj, 342, L87
\bibitem{} Sunada, K., Yamaguchi, C., Nakai, N., et al. 2000, in Proc. SPIE
Vol.4015, Radio Telescopes, ed. H.R. Butcher, 237
\bibitem{} Tej, A., Ojha, D. K., Ghosh, S. K. et al. 2006, A\&A, in press
\bibitem{} Torrelles, J. M., Verdes-Montenegro, L., \& Ho, P. T. P. 1993, \apj, 410, 202
\bibitem{} Turner, B. E., \& Matthews, H. E. 1984, \apj, 277, 164
\bibitem{} van der Walt, D. J., Churchwell, E., Gaylard, M. J., \& Goedhart, S. 2003, MNRAS, 341, 270
\bibitem{} Verdes-Montenegro, L., Torrelles, J. M., Rodriguez, L. F. et al. 1989, \apj, 346, 193
\bibitem{} Walker, R. N. F., \& Masheder, M. R. X. 1997, \mnras, 258, 862
\bibitem{} Walsh, A. J., Burton, M. G., Hyland, A. R., \& Robinson, G. 1998, MNRAS, 301, 640
\bibitem{} Wilking, B. A., Blackwell, J. H., Mundy, L. G., \& Howe, J. E., 1989, \apj, 345, 257
\bibitem{} Xu, Y., Zheng X. W., \& Jiang, D. R. 2003, ChJAA, 3, 1663
\bibitem{} Xu, Y., Yang J., Zheng, X. W. et al. 2004, Chin. Phys. Letter., 21, 2071
\bibitem{} Yang, J., Jiang, Z., Wang, M., Ju, B., \& Wang, H. 2002, \apjs, 141, 157
\bibitem{} Yang, J., Umemoto, T., Iwata, T., \& Fukui, Y. 1991, \apj, 373, 137
\bibitem{} Yao, Y., Hirata, N., Ishii, M. et al. 1997, \apj, 490, 281
\bibitem{} Yao, Y., Ishii, M., Nagata, T. et al. 2000, \apj, 542, 392
\bibitem{} Zhang, Q., Hunter, T. R., Brand, J. et al. 2001, \apj, 552, L167
\bibitem{} Zhang, Q., Hunter, T. R., Brand, J. et al. 2005, \apj, 625, 864
\bibitem{} Zheng, X. W., Ho, P. T. P., Reid, M. J., \& Schneps, M. H. 1985, \apj, 293, 522
\bibitem{} Zinchenko, I., Henning, Th., \& Schreyer, K. 1997,
A\&AS, 124, 385
\bibitem{} Zinchenko, I., Pirogov, L., \& Toriseva, M. 1998,
A\&AS, 133, 337
\bibitem{} Zuckerman, M, Kuiper, T. B. H., \& Rodriguez Kuiper, E. N. 1976, \apj, 209,
L137 }
\end{thebibliography}
\end{document}